\documentclass{elsart}
\usepackage{epsfig}
\usepackage{rotate}
\usepackage{amssymb}
\usepackage{amsmath}

\newcommand{\eg}{{\it{e.g.}}}
\newcommand{\ie}{{\it{i.e.}}}
\newcommand{\ps}{\ensuremath{\mathrm{Ps}}}
\newcommand{\ops}{\ensuremath{\mathrm{o\text{-}Ps}}}
\newcommand{\decay}{\ensuremath{\ops\to \gamma +\mathrm{ X_1 + X_2}}}
\newcommand{\bdecay}{\ensuremath{B(\decay)}}
\newcommand{\xone}{\ensuremath{\mathrm{X_1}}}
\newcommand{\xtwo}{\ensuremath{\mathrm{X_2}}}
\newcommand{\me}{\ensuremath{m_{\mathrm{e}}}}
\newcommand{\mone}{\ensuremath{m_\xone}}
\newcommand{\mtwo}{\ensuremath{m_\xtwo}}

\newcommand{\eveto}{\ensuremath{E_{\mathrm VETO}}} 
\newcommand{\etrig}{\ensuremath{E_{\mathrm TRIG}}}

\newcommand{\tibgo}{\Delta t_{\mathrm{e^+}\gamma}}

\begin{document}
\begin{frontmatter}
\title{\boldmath Search for an exotic three-body decay of orthopositronium}
\vspace{.5cm}
\author[Zuerich]             {A.~Badertscher},
\author[Zuerich]             {P.~Crivelli},
\author[Zuerich]              {M.~Felcini},
\author[INR]               {S.N.~Gninenko},
\author[INR]               {N.A.~Goloubev},
\author[LAPP]              {P.~N\'ed\'elec},
\author[LAPP]            {J.P.~Peigneux},
\author[INR]               {V.~Postoev},
\author[Zuerich]      {A.~Rubbia},
\author[LAPP]              {D.~Sillou}

\address[Zuerich]        {ETH Z\"urich, Z\"urich, Switzerland}
\address[INR]            {Inst. Nucl. Research, INR Moscow, Russia}
\address[LAPP]           {CNRS-IN2P3, France}

\begin{abstract}
We report on a direct search  for a three-body decay of the 
orthopositronium 
into a photon and two penetrating
particles, \decay. The existence of this decay could  explain the discrepancy
between the measured and the predicted values of the 
orthopositronium decay rate.
From the analysis of the collected data  a single candidate event is found,
consistent with the expected background. 
This allows to set an upper limit on the branching ratio
\bdecay$< 4.4 \times 10^{-5}$ (at the 90\% confidence level), 
for the photon energy in the range 
from 40 keV $< E_{\gamma}<$ 400 keV and 
for mass values in 
the kinematical range  $0\leq\mone+\mtwo\leq$ 900 keV.
This result unambiguously
excludes the \decay\ decay mode as the origin 
of the discrepancy.
\end{abstract}
\begin{keyword} 
orthopositronium decay, new particles, experimental tests
\end{keyword}
\end{frontmatter}

\section{Introduction}

Positronium (\ps ), the positron-electron bound state,
is the lightest known atomic system. It is bound and annihilates 
through the electromagnetic interaction alone. At the 
current level of experimental and theoretical precision this is  
the only interaction present in this system (see \eg\ \cite{rubbia}). 
This feature has made the positronium an ideal system for  
testing the accuracy of the QED calculations 
for bound states, in particular for the triplet ($1^3S_1$)
state of \ps, called orthopositronium (\ops). 
Due to the odd-parity under C-transformation  the \ops\ decays
predominantly into three photons. 
As compared with the singlet ($1^1S_0$) state (parapositronium) decay rate,
the \ops\ decay rate, due to the phase-space and  additional
$\alpha$ suppression factors, is about $ 10^3$ times smaller.
Thus, it is  more sensitive to potential  admixtures of 
new interactions, which are not accommodated in the Standard Model.
 
The study of the \ops\ system has a long history \cite{rich}.  However,
in spite of the substantial efforts devoted to the
theoretical and experimental determination of the \ops\ properties,  
there is a long-standing  puzzle: the \ops\ decay rate in vacuum 
measured by the Ann Arbor group, 
$\Gamma^{exp}=7.0482\pm 0.0016 ~\mu s^{-1}$ \cite{nico}, has a 
$\simeq 5\sigma$ discrepancy with respect to the  
predicted value $\Gamma =7.03830\pm 0.00007 ~\mu s^{-1}$
\cite{adkins1} (see also \cite{lepage}). 
This discrepancy has been recently confirmed by more precise calculations
of Adkins \etal \cite{adkins2}, 
including corrections of the order $\alpha^2$.

The result of the recent Tokyo measurements of \ops\ decay rate
in low density SiO$_2$ powder corrected for matter effects \cite{asai}
agrees, within the errors,  with the theoretical value of \cite{adkins2}.
However, the method to extract the value of the \ops\ decay rate in vacuum 
from the  measurement in matter is still under discussion \cite{scals}. 
Thus, it is difficult to disagree with the Adkins \etal\cite{adkins2} statement that : {\em...no conclusions can be 
drawn until the experimental situation is clarified}.

Various exotic \ops\ decay modes have been investigated, with the hope 
that a relative contribution to the \ops\ decay rate at the level 
of $\Delta\Gamma=(\Gamma^{exp}- \Gamma)/\Gamma \simeq 10^{-3}$ 
would solve the discrepancy\footnote{Note that the recent result \cite{asai}
 still allows exotic 
contribution to the \ops\ at a (2$\sigma$) level of $\Delta \Gamma 
\simeq 7\times10^{-4}$.}
(for review, see \eg\ \cite{scals,dobr}).

Invisible decays of \ops, such as, \eg, $\ops\to \nu \overline{\nu}$,
are excluded \cite{atojan,mits}. However, there is still 
a possible explanation of both the discrepancy 
and the Tokyo results. This is  based on the existence of the
$\ops \to \text{invisible particles}$ 
decay in {\em vacuum} (for more details see 
Ref. \cite{glashow}--\cite{foot}).

Visible exotic  decays of \ops\ (\ie\ decays accompanied by at least 
one photon in the final state) have been experimentally searched for:
$\ops\to \gamma+\mathrm{X}$, 
$\ops\to \gamma \gamma \mathrm{X}$  and $\ops\rightarrow N \gamma$,
where  X is a new light particle and $N=2,4,..$. 
These decay modes have definitely been excluded \cite{klubak}--\cite{mitsui}
as the cause of the \ops\ lifetime discrepancy. 

The possibility  of a new exotic three-body decay \decay\ 
into a photon and a pair of new light weakly interacting particles 
had not yet been considered.
The photon energy spectrum from this decay mode has no peak, 
differently from the two-body decay $\ops\to \gamma+\mathrm{X}$.  
Thus, the sensitivity of  previous searches based on the conventional 
peak-hunting technique is not sufficient to exclude this decay mode as the 
source of the discrepancy.
Cosmological arguments place an indirect stringent 
constraint on the branching ratio of this decay mode \cite{escri}. 
However, one may argue that these arguments are depending on many assumptions.

The purpose of this experiment is to perform a direct 
search for the decay \decay\ with the sensitivity needed to 
exclude this decay unambiguously.

 

\section{Experimental technique}

\begin{figure}[htb]
\hspace{.0cm}\includegraphics[height=.25\textheight,width=.5\textwidth]{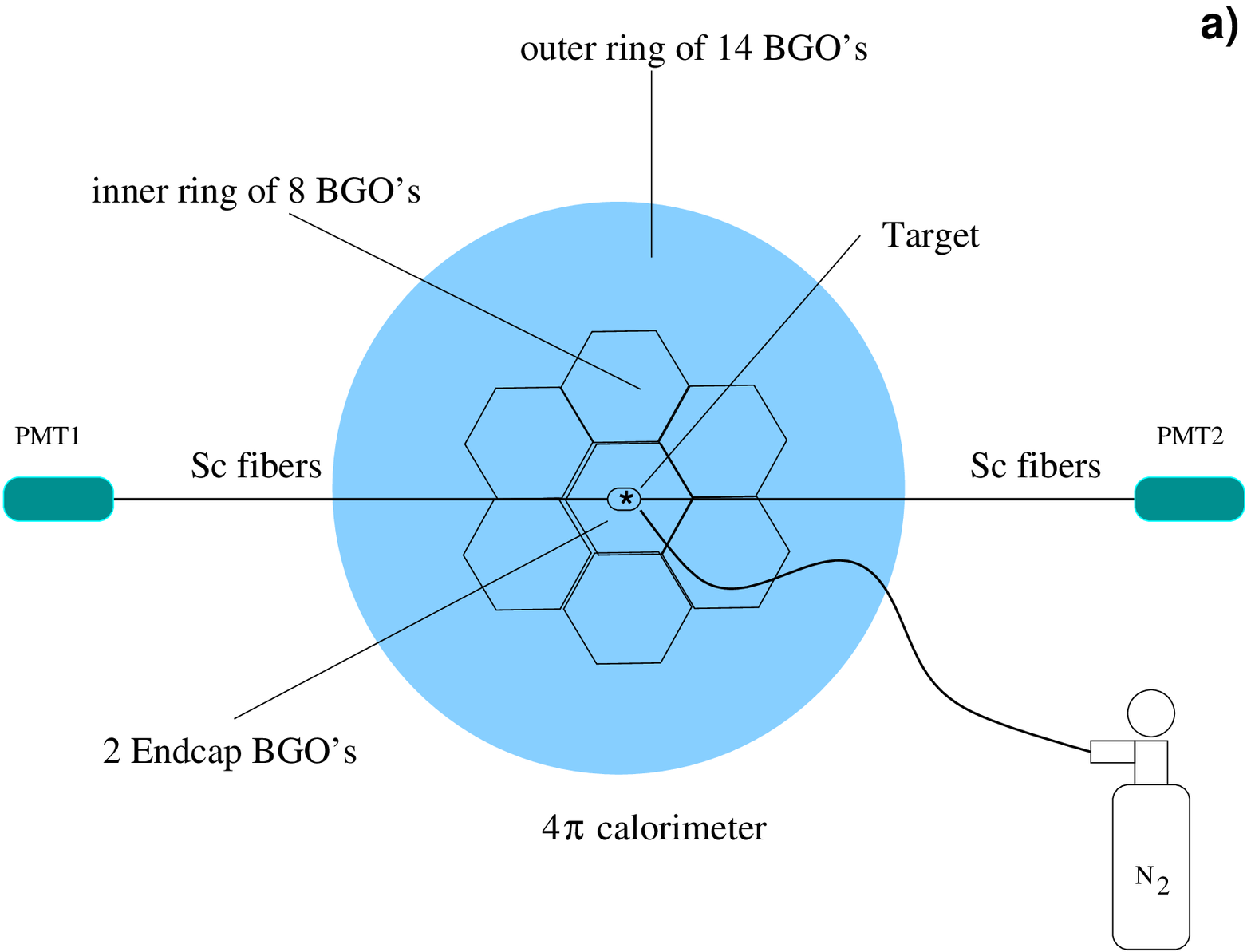}
\hspace{.0cm}\includegraphics[height=.25\textheight,width=.5\textwidth]{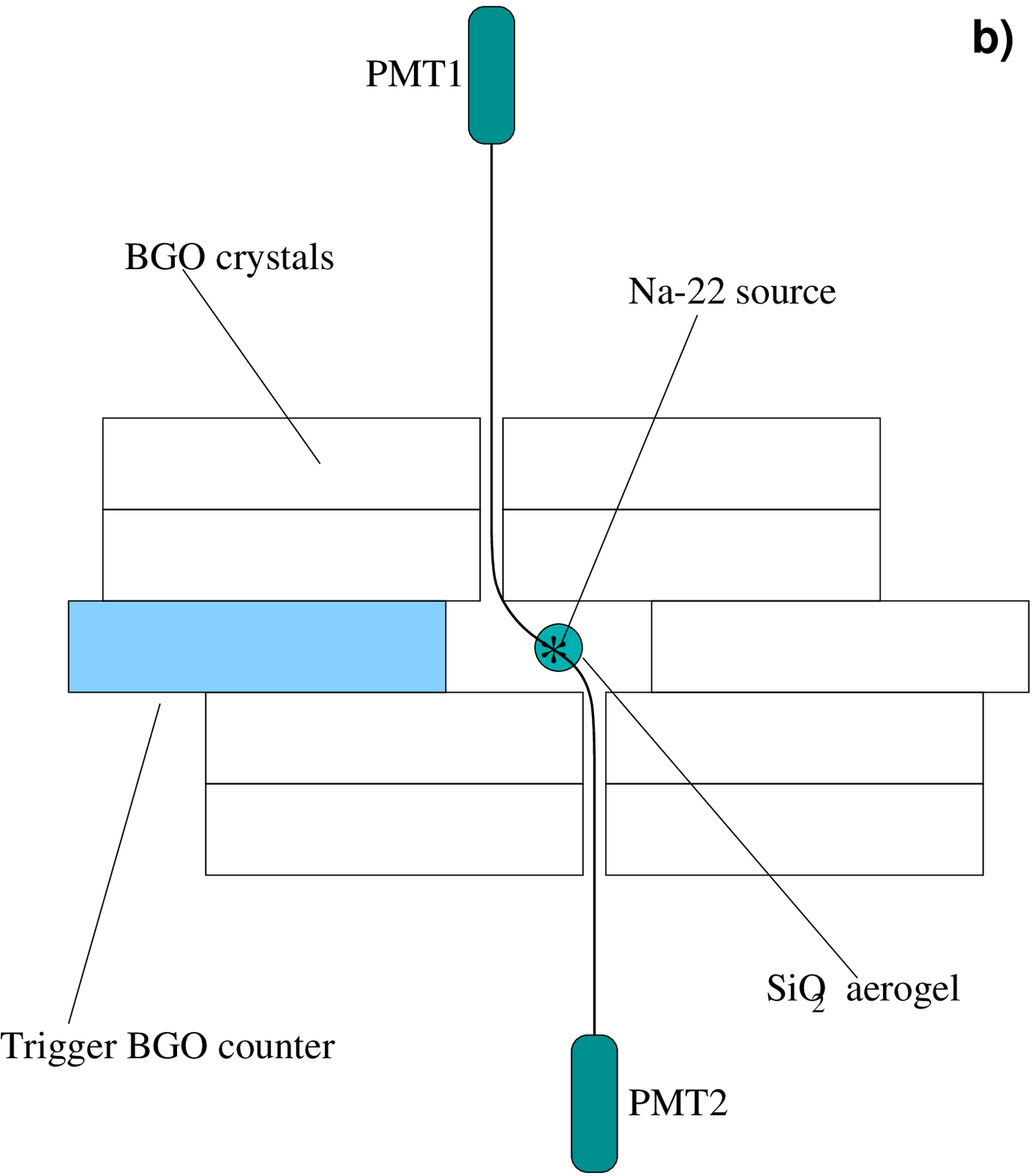}
\vspace*{-.cm}
\caption{\em Schematic illustration of the experimental setup: a) front view, b) top view.}
\label{detector}
\end{figure}

The schematic illustration of the detector setup used in the experiment 
is shown in Figure \ref{detector}.  Positrons from the $^{22}$Na source 
with an activity of 3.6 kBq  are stopped in the SiO$_2$ aerogel
target (density $\rho \simeq$ 0.1 g/cm$^3$, average grain size 50 - 100 {\AA}) 
where a fraction of them produces orthopositronium. The source is prepared 
by sealing a drop of the  $^{22}$Na solution between two 
5 $\mu$m mylar foils. The  source     
is sandwiched between two 120 $\mu$m thick scintillators fabricated by
squeezing 1 mm thick scintillator fibers.
The light produced by positrons crossing one of the scintillator fibers
is delivered by the fibers to a pair of photomultipliers
Philips XP2020 (PMT1 and PMT2 in Figure \ref{detector}). 
The coincidence of the signals from PMT1 and PMT2 is used to
tag the positron emission ($e^+$ trigger) and to define the time $t_0$ 
of positronium formation in the target.  
 
The photons produced by the positronium annihilation are detected 
by a 4$\pi$ crystal calorimeter, schematically shown in Figure \ref{detector}.  
The detector is composed of an inner and an outer ring
of 8 and 14 BGO crystals, respectively,
surrounding the target region. 
Two additional BGO crystals serve as endcaps.
Each crystal has a hexagonal cross-section with an inner diameter
  of 55 mm and a length of 200 mm. For a more detailed description of the 
  BGO crystals see \cite{bgo}.
One of the endcap counters, hereafter called trigger counter, 
is also used for the measurement of the annihilation time 
of the positronium relative to $t_0$.   
When one of the annihilation photons 
is detected by the trigger counter, 
the recoil photons are detected by the other BGO counters,
hereafter called the VETO detector.
The detector is calibrated and monitored internally 
using  the 511 keV annihilation photons  
and the 1.27 MeV photon emitted by the $^{22}$Na source 
in association  with the positron emission.
Variations of the energy scale are within $\lesssim 1\%$ 
and are corrected on the basis of an internal calibration procedure.

The particles \xone\ and \xtwo\  are assumed to be  weakly interacting 
penetrating particles.
Thus the experimental signature of the \decay\ decay 
is the presence of energy deposition  
in the trigger counter, within a time interval consistent with the 
 delayed annihilation  of the \ops\  in the target, 
and no energy deposition in the VETO detector.

 In order to decrease the contribution of 
two photon events from collisional quenching of \ops,
the SiO$_2$ target is dehydrated in a vacuum of 
10$^{-2}$ Torr at a temperature about 200$^o$C during two hours
before installation inside the detector. Furthermore,
during the data taking period  high purity dry nitrogen is flowing 
through the target. This procedure increases the lifetime of \ops\ 
in the prepared sample of the SiO$_2$ aerogel from $\simeq$ 70 ns   to 132 ns  
giving a two photon  suppression factor of $\simeq 10$.

The signal of positron emission is also used to open 
a 3 $\mu$s gate for complete
recording of the signals from the BGO counters and the scintillator PMTs. 
An event is recorded if an energy deposition $\etrig\geq$ 40 keV  
is detected in the trigger counter within the 3 $\mu$s gate.
 
A CAMAC-VME system interfaced to a personal computer is used for data
acquisition. For each event the following quantities are
recorded:
\begin{itemize}
\item the amplitudes, $A_1$ and $A_2$,  of the pulses from PMT1 and PMT2   
and the time interval $\Delta t_{12}$ between them;
\item the energy deposition 
\etrig\ in the trigger counter and the  
time interval $\Delta t_{e^+\gamma}$ between the trigger counter pulse and 
the $e^+$ trigger;
\item the pulses from each of the 23 BGO crystals for the measurement of the 
total energy deposition in the VETO detector.
  
\end{itemize}


\section{Results}  

\begin{figure}[htb]
\mbox{\epsfig{figure=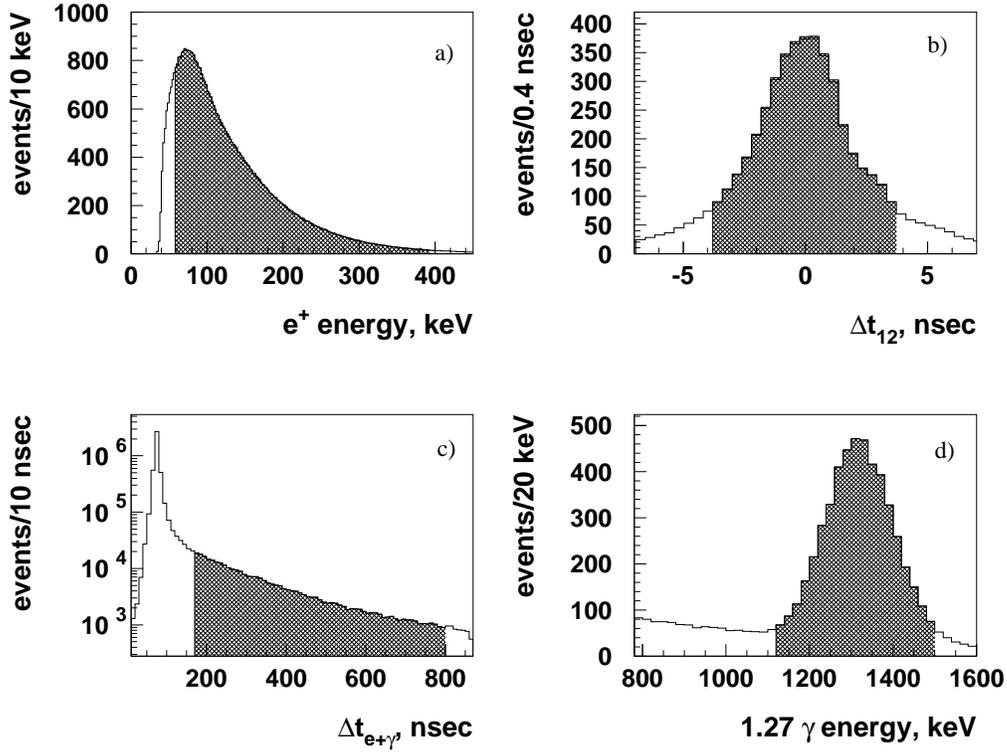,width=1.0\textwidth}}
\caption{\em a) Distribution of the energy deposited by the positrons in the 
scintillator fibers; 
b) Time difference between the two photomultipliers pulses from the 
scintillator fibers; c) Time difference between the scintillator 
fiber pulses and the trigger counter; 
d) 1.27 MeV photon energy spectrum in the 
BGO crystals. The dashed areas  on the plots illustrate the corresponding cuts 
used for the event selection. See Section 3 for details.}
\label{cuts}
\end{figure}

The search for the \decay\ decay described in this paper uses a data 
sample  of $2\times 10^{7}$ recorded events. 
Candidate events for o-Ps decays, 
with  at least one photon in the final state,  
are identified by the following  selection criteria,
 illustrated in Figure \ref{cuts}:  
\begin{itemize}

\item the PMT1 and PMT2 pulses from the scintillator fibers 
in the energy range 
 $60 \text{ keV}<A_{1,2}<400$ keV. This cut is applied to reject signals 
of fake positrons from accidental coincidences due to PMT noise;

\item the time difference between PMT1 and PMT2 pulses 
$|\Delta t_{12}|<3.8$ ns;

\item the time difference between the PMT pulses from the scintillator fibers 
and the trigger counter signal is required to be in the range 
$160 <\tibgo< 800$ ns. The lower cut is used to 
eliminate background from the tail of prompt positron annihilation, shown
in Figure 2c). The upper cut is chosen  
to eliminate a region where the background
 from accidental coincidences becomes dominant.

\item the presence in any single BGO crystal, except the two endcaps,
of an energy deposition in the range $1100\text{ keV} < E_{1.27} < 1500$ keV. 
This criterion is applied to select events with a
1.27 MeV photon, emitted by the source at the same time
as the positron.
 
\end{itemize}

After imposing the above requirements  338'786 candidates events 
are found. 
The distribution of these events 
in the scatter plot \eveto\ vs \etrig\ is shown in Figure 3a). 
Here \eveto\ is defined as 
the sum over all BGO crystals minus the energy deposited in the trigger
crystal and  the energy deposited
 by the 1.27 MeV photon: 
$\eveto=\Sigma E_i -\etrig\ -E_{1.27}$.
As expected for positronium decay into photons, 
the events accumulate around the line \eveto+\etrig=2\me,
\me=511 keV being the electron or positron mass.

In Figure 3b) the region of the scatter plot for \eveto$<35$ keV 
is shown. The box for $\eveto<$ 20 keV and  $\etrig<$ 400 keV
defines the signal region. 
The cut on the trigger energy $\etrig < $ 400 keV is applied to 
reject events where two photons from the $\ops \to 3 \gamma$ decay
deposit energy in the  trigger counter
and to eliminate background from \ops\ annihilation into two photons
due to collisional quenching. 

The cut $\eveto<$ 20 keV
is chosen according to the measured width of the zero-energy peak
in the VETO detector.
For this measurement, prompt positron annihilation
events ($\tibgo< 160$ ns) are selected with one 
511 keV photon in the trigger counter. 
The spectrum measured in the VETO counter 
for zero-energy signal, when the second photon escapes detection,
shows that 95\% of the zero-energy signal is collected in the region 
$\eveto < 20$ keV. The width of the zero-energy signal spectrum  
is determined mostly by the  overlap of close in time events, 
while the contribution from the ADC pedestals fluctuation is found to be 
negligible. 

\begin{figure}[htb]
\mbox{\epsfig{figure=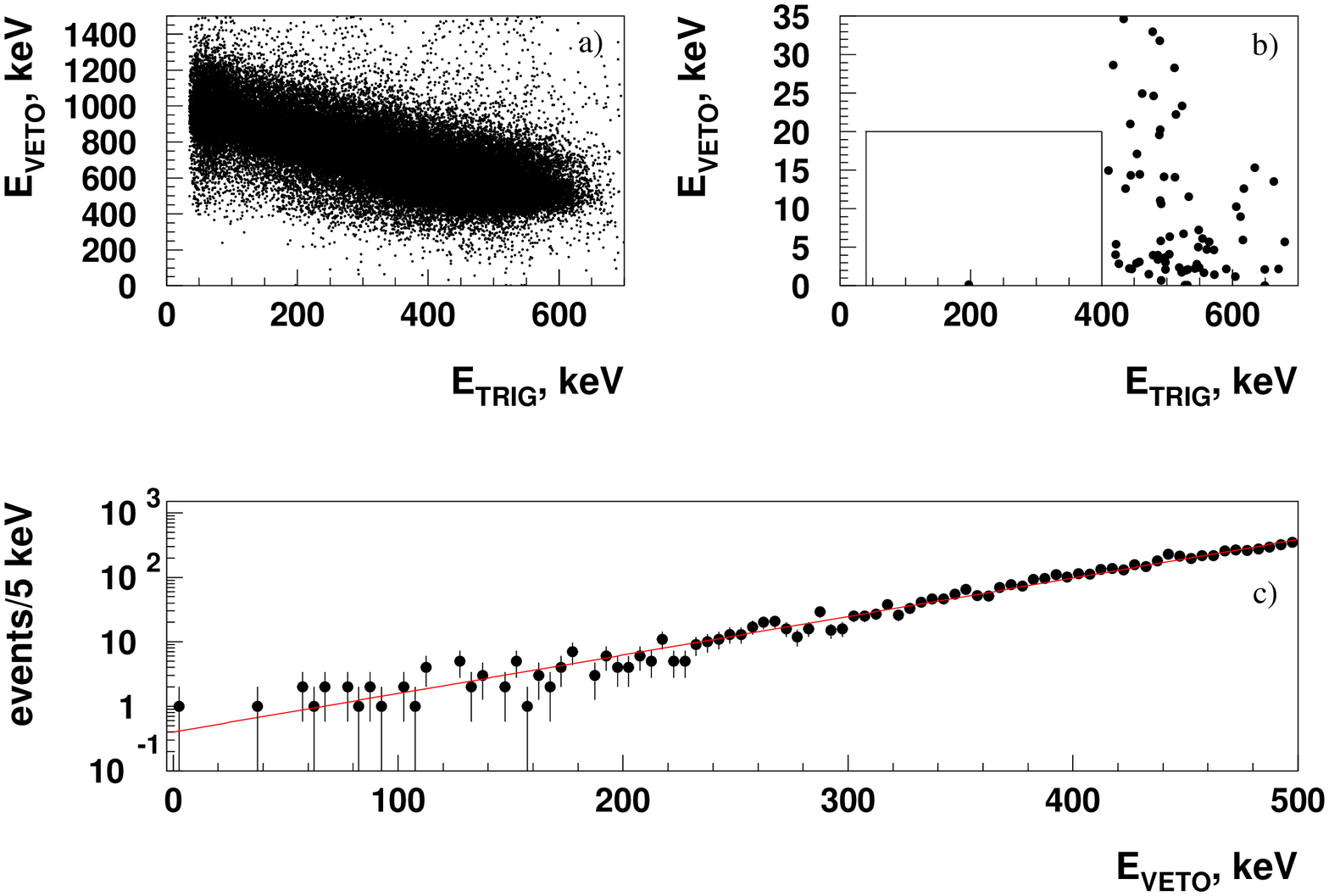,width=1.0\textwidth}}\vspace*{-0.5cm}
\caption{\em Distributions  for the selected \ops\ events: 
a) distribution of the VETO energy \eveto\ vs the  energy 
\etrig\ in the trigger counter; 
b) close-up of the scatter plot for $\eveto<35$ keV;
the rectangle  represents the signal region for $\decay$ events; 
c) distribution of \eveto\ for events in the region 
$40\text{ keV} < \etrig < 400$ keV.
The line represent the result of a fit of the \eveto\ distribution, allowing
to estimate the background contribution in the signal region. 
One event is observed in the signal region. 
\vspace*{0.5cm}
}
\label{result}
\end{figure}

The distribution of the VETO energy  for $\eveto <$ 500 keV
is shown in Figure \ref{result}c). 
The distribution of the events with $\eveto > 20$ keV 
can be extrapolated into the signal region to evalute the background
contribution in this region. 
A fit  shown in Figure \ref{result}c) results in a background estimate of
$N_{bckg}=1.6\pm 0.8$ events, where 
the error is evaluated from the uncertainty related to the 
extrapolation procedure itself. 

As shown in  Figure \ref{result}b) and \ref{result}c)
one event is found in the signal region. 
This is consistent with the background evaluation described above. 
Hence, no evidence for the decay \decay\ is found.

This result allows us  to set an 
upper limit on \bdecay\ 
from the 90\% confidence level (CL) 
upper limit on the expected number of signal events,
$N_{\decay}^{\rm up}$.  Because of the uncertainty on the background 
estimate we have chosen  conservatively not to subtract the background. 
Using Poisson statistics \cite{pdg}, 
for 1 event observed and 0  background event expected, the limit  
 is $N_{\decay}^{\rm up}$=3.8 events.   

Given the measured number $N_{\ops}$ of \ops\ decays in the target, 
the numbers $N_{\decay}$ and $N_{\ops\to 3 \gamma}$ of \decay\  and 
 $\ops\to 3 \gamma$ decays , respectively,
detected in the trigger counter,  are  
\begin{eqnarray}
N_{\decay}= N_{\ops} BR(\decay) \epsilon_{1\gamma}\ , \\
N_{\ops\to 3 \gamma} = N_{\ops} \epsilon_{3\gamma}\ , 
\end{eqnarray}
resulting in 
\begin{equation}
Br(\decay) < \frac{\epsilon_{3\gamma}}{\epsilon_{1\gamma}}\frac{N_{\decay}^{\rm up}}{N_{\ops\to 3\gamma}\ .}
\end{equation}
Here
$\epsilon_{1\gamma}$ and $\epsilon_{3\gamma}$ are the  
efficiencies  of the trigger counter for  \decay\ and for $\ops\to 3\gamma$
decays, respectively. These efficiencies are 
evaluated with a Monte Carlo simulation of the detector response to 
both decay modes. For the \decay\ decay,  
phase space  distribution of the photon and of the two particles \xone\ and 
\xtwo\ are considered. It is found that 
for the kinematically allowed
region of masses $0\leq\mone+\mtwo\leq 900$ keV 
the efficiency ratio varies in the range  
$3.0<\epsilon_{3\gamma}/\epsilon_{1\gamma}<3.7$.
The lower value corresponds to $\epsilon_{1\gamma}$ 
calculated for $\mone+\mtwo$=0 keV. The upper value corresponds to   
$\mone+\mtwo$=900 keV. The efficiency for the 
\decay\ is relatively high over this
kinematically allowed region owing to  
our low trigger threshold of 40 keV.  
For the determination of the limit we take conservatively
the value $\epsilon_{3\gamma}/\epsilon_{1\gamma}=3.7$.   

The number of \ops\ decays in the target  is measured 
from the decay curve of Figure 2c by fitting the distribution to the 
function $A\cdot exp(-t/\tau_{\ops}) + B$ ($B$ is the accidental background)
starting from the time $t=160$ ns when \ops\ is completely 
thermalized in the target.
Comparing the measured lifetime $\tau_{\ops}=132.5\pm3.2$ ns with the 
lifetime in vacuum ( 141.9 ns) the probability of \ops\ quenching in 
the target is found to be 6.6\%. Correcting for this efficiency factor
the total number of detected \ops\  is determined to 
be 3.2$\times 10^5$. 

Finally, our 90 \% CL limit  on \bdecay\ for the photon energy range 
  $40\text{ keV} < E_{\gamma}< 400$ keV and masses 
$ m_{X_1}+m_{X_2} \leq 900$ keV is

\begin{equation}
\bdecay < 4.4 \times 10^{-5} 
\end{equation}

This limit is more than 20 times smaller than the value needed to explain the 
discrepancy in \ops\ decay rate. Thus, the \decay\ decay is 
definitely excluded as a possible origin of the discrepancy.

\section*{Acknowledgments}
We thank the Paul Scherrer Institute (PSI)
for providing us with the BGO crystals 
and W. Fetscher for advice.  We gratefully acknowledge the help of
G.~Roubaud for the $^{22}$Na source preparation. We wish to thank 
N.V. Krasnikov for useful  discussions.\\ 
Part of this work was supported by the Swiss National Science Foundation.

\end{document}